\documentstyle[prl ,aps, graphics, epsfig, floats]{revtex}
\twocolumn

\tighten
\draft

\begin{document}

\title{Quantum communication complexity protocol with two entangled qutrits}

\author{{\v C}aslav Brukner$^1$, Marek Zukowski$^2$, Anton Zeilinger$^1$}
\address{$^1$Institut f\"ur Experimentalphysik, Universit\"at Wien,  Boltzmanngasse 5, A--1090 \\
$^2$Instytut Fizyki Teoretycznej i Astrofizyki Uniwersytet
Gda\'nski, PL-80-952 Gda\'nsk, Poland}

\date{\today}
\maketitle

\begin{abstract}

We formulate a two-party communication complexity problem and
present its quantum solution that exploits the entanglement
between two qutrits. We prove that for a broad class of protocols
the entangled state can enhance the efficiency of solving the problem
in the quantum protocol over any classical one {\it if and only if}
the state violates Bell's inequality for two qutrits.

\end{abstract}


\vspace{-0.5cm}

Entanglement is not only the most distinctive feature of quantum
physics with respect to the classical world, as quantitatively expressed by
the violation of the Bell's inequalities \cite{bell}. It also 
enables powerful computation \cite{nielsen}, establishes secure
communication \cite{nielsen} and reduces the communication
complexity \cite{yao,brassard,cleve,grover1,buhrman,buhrman1} all beyond
the limits that are achievable on the basis of laws of classical physics.

To date only very few tasks in quantum communication and quantum
computation require higher-dimensional systems than qubits as
recourses. Quantum-key distribution based on higher alphabets was
shown to be more secure than that based on qubits \cite{bruss}.
A certain quantum solution of the coin-flipping problem uses
qutrits (3-dimensional quantum states) \cite{ambainis} and the
quantum solution of the Byzantine agreement problem utilizes
the entanglement between three qutrits \cite{gisin}.

Here we formulate a two-party communication complexity
problem and present its quantum solution which makes use of the
entanglement between {\it two} qutrits. We prove that for a broad class of protocols
the entangled state of two qutrits can enhance the efficiency of solving the
problem in the quantum protocol, over any classical one {\it if and only if} the state
violates Bell's inequality for two qutrits as derived by
Collins {\it et al.} \cite{collins}.

In this paper a variation of the following communication complexity problem will be
considered. Two separated parties (Alice and Bob)
receive some input data of which they know only their own data and not the
data of the partner. Alice receives 
an input string $x$ and Bob an input string $y$ and the
goal is for both of them to determine the value of a certain
function $f(x,y)$, while exchanging a {\it restricted} amount of
information. While an error in
computing the function is allowed, the parties try to compute it
correctly with as high probability as possible. An
execution is considered successful if the value determined by
both parties is correct. Before they start the protocol
Alice and Bob are allowed to share (classically correlated) random strings which 
might improve the probability of success.

In 1997 Buhrman, Cleve and van Dam \cite{buhrman} considered a
specific two-party communication complexity problem of the type
given above. Alice receives a string $x = (x_0,x_1) $ and Bob
a string $y = (y_0,y_1) $. Each of the strings is a combination
of two bit values: $x_0, y_0  \in
\{0,1\}$ and $x_1, y_1 \in \{ -1,1\}$. Their common goal is to
compute the function (a reformulation of the original
function of \cite{buhrman})
\begin{equation}
f(x,y) = x_1 \cdot y_1 \cdot (-1)^{x_0\cdot y_0}
\label{prva}
\end{equation}
with as high probability as possible, while exchanging altogether only 2 bits
of information. Buhrman {\it et al.} \cite{buhrman}
showed that this can be done with a probability of success
of $P_Q= 0.85$ if the two parties share two qubits in a maximally
entangled state, whereas with shared random
variables but without entanglement (i.e. in a classical protocol) this probability
cannot exceed $P_C=0.75$. Therefore in a classical protocol 3 bits of information
are {\it necessary} to compute $f$ with a probability of at least 0.85,
whereas with the use of entanglement 2 bits of information are {\it
sufficient} to compute $f$ with the same probability.

There is a link between tests of Bell's inequalities and quantum
communication complexity protocols. Bell's inequalities are bounds
on certain combinations of probabilities or correlation
functions for measurements on multi-particle systems. These bounds apply
for any local realistic theory. In a realistic
theory the measurement results are determined by 
properties the particles carry. In a local theory the results obtained at
one location are independent of any actions
performed at space-like separation. The quantum protocol of the two-party communication complexity 
problem introduced in Ref. \cite{buhrman,buhrman1}
is based on a violation of the Clauser-Horne-Shimony-Holt \cite{chsh} inequality.
Similarly three- and multi-party communication complexity
tasks were introduced \cite{buhrman,buhrman1,galvao} with quantum solutions
based on the GHZ-type \cite{ghz} argument against local realism.

Let us now define the two-party communication complexity
problem which will be our case of study. 
This problem is of a different kind than the standard communication complexity problem where the
two parties try to give the correct answer to a question posed to
them in as many cases as possible under the constraint of restricted
communication. Yet, one can imagine situations where not a single but two
questions are posed to the parties and where the parties are restricted both in communication and in
broadcasting of their answers. The specific case which is considered here is that  
the parties must give a {\it single} answer to {\it two} questions. Further 
the parties are {\it not} allowed to differ in their answers. 
That is, they {\it must} produce two identical answers each time.

Formally the two questions will be formulated as a problem of computation of
two three-valued functions $f_1$ and $f_2$. Since the parties are allowed to give 
only one answer about the values of two functions, their goal will be to give
the correct value of $f_1$ with the {\it highest} possible probability, and at
the same time, the correct value of $f_2$ with the {\it lowest} possible probability,
while exchanging only a restricted (each party sending one trit) amount of information \cite{MOTIVATION}.

\begin{figure}
\hspace{-0.3cm} \centerline{\psfig{width=4cm,file=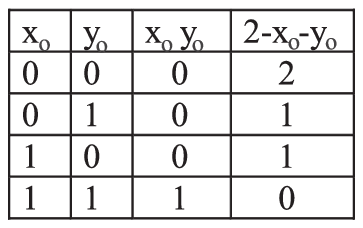}}
\vspace{-0.2cm} \caption{A set of possible input values for $x_0$
and $y_0$ and the corresponding values of the functions $x_0
\!\cdot\! y_0$ and $2\!-\!x_0\!-\!y_0$.} \label{table1}
\vspace{-0.5cm}
\end{figure}

We now introduce the two-party task in detail and we give the functions $f_1$ and $f_2$ explicitly:
\begin{itemize}
\item Alice receives a string $x = (x_0,x_1) $ and Bob a string
$y = (y_0,y_1) $. Alice's string is a combination of a bit $x_0 \in \{0,1\}$ and a trit
$x_1 \in \{ 1, e^{i\frac{2\pi}{3}}, e^{-i\frac{2\pi}{3}} \}$.
Similarly Bob's string is a combination of a bit $y_0 \in \{0,1\}$ and a trit
$y_1 \in \{ 1, e^{i\frac{2\pi}{3}},
e^{-i\frac{2\pi}{3}} \}$ (the representation in terms of complex 3-rd roots of unity is chosen
for mathematical convenience). All possible input strings are distributed randomly and with equal probability. 
\item Before they broadcast their answers Alice and Bob are allowed to exchange 2 trits of information.
\item Alice and Bob each broadcast her/his answer in the form of one trit. The two answers must be identical. That is, they each broadcast the same one trit. 
\item The task of Alice and Bob is to maximize the difference between the probabilities, $P(f_1)$, of giving the correct value for function
\begin{equation}
f_1 = x_1 \cdot y_1 \cdot e^{i\frac{2\pi}{3} (x_0 \cdot y_0)},
\label{f1}
\end{equation}
and $P(f_2)$, of giving the correct value for function
\begin{equation}
f_2 = x_1 \cdot y_1 \cdot e^{i\frac{2\pi}{3} (2-x_0 -y_0)}.
\label{f2}
\end{equation}
That is, they aim at the maximal value of 
\begin{equation}
\Delta=P(f_1)-P(f_2). 
\end{equation}
\end{itemize}

We will show that if two parties use a broad class of classical
protocols the difference $\Delta$ of the probabilities for correct
value of the two functions introduced above is at most 0.5, whereas,
if they use two entangled qutrits this difference can be as large
as $1/4 + 1/4\sqrt{11/3} \simeq 0.729$.

Note, that the first factor $x_1 \cdot y_1$ in the full functions
$f_1$ and $f_2$ results in completely random values if only one of the
independent inputs $x_1$ or $y_1$ is random. This is not the case
for the last factors with the inputs $x_0$ and $y_0$. Thus {\it intuition}
suggests that the optimal protocol for the two parties may be that Alice "spends" her trit in sending
$x_1$ and Bob in sending $y_1$ and that they put for the second factor of the two functions a value
which is most often appearing for function $e^{i\frac{2\pi}{3} (x_0 \cdot y_0)}$ (compare the third column in table 1)
and, at the same time, least often appearing for function $e^{i\frac{2\pi}{3} (2-x_0 -y_0)}$ (compare the fourth
column). The second factor obtained in such a way is 1. Next, each of them broadcasts
the value $x_1\cdot y_1$ as her/his answer. In this way $P(f_1)\!=\!0.75$, and $P(f_2)\!=\!0.25$, which gives $\Delta\! =\! 0.5$.

The second protocol suggested by intuition exploits the fact that $f_2$, in contradiction to $f_1$, is a
factorizable function, 
i.e. $f_2=(x_1\cdot e^{i\frac{2\pi}{3} (1-x_0)} )\cdot( y_1 \cdot e^{i\frac{2\pi}{3} (1-y_0)}).$ 
Alice and Bob can exchange the values of $x_1\cdot e^{i\frac{2\pi}{3} (1-x_0)}$  and 
$ y_1 \cdot e^{i\frac{2\pi}{3} (1-y_0)}$. In this way they both know the exact value of $f_2$. 
Thus they broadcast a wrong value of it, e.g. $f_2\cdot e^{-i\frac{2\pi}{3}}$. By looking 
at the table \ref{table1} one immediately sees that this operation acts effectively as
subtraction of $1$ from the values in the last column. The obtained values agree with those in column
three in two cases (two middle rows). Thus within this protocol 
$P(f_1)=0.5$, and $P(f_2)=0$, which again results in $\Delta = 0.5$. 

Let us now present a broad class of classical protocols which can be followed by Alice and Bob, 
and which contain the above intuitive examples as special cases:
\begin{itemize}
\item
Alice calculates locally any function $a(x_0, \lambda_A)$ and Bob calculates locally any function
$b(y_0, \lambda_B)$ such that their outputs define the trit values to be broadcast under 
the restriction of communication. More precisely, Alice sends to Bob $e_A\!=\!a\cdot x_1$ and
receives from him $e_B\!=\!b\cdot y_1$. Here $\lambda_A$ and $\lambda_B$ are any other
parameters on which their functions $a$ and $b$ may depend. They may include random strings of numbers
shared by Alice and Bob. 
\item Upon receipt of $e_A$ and $e_B$ they both broadcast $e_A\cdot e_B$ as their
answers (which always agree).
\end{itemize}
Note that the first intuitive protocol is reproduced by $a=1$ and $b=1$ for all inputs.
The second one is recovered by $a= e^{i\frac{2\pi}{3}(1-x_0)}$  and 
$b=e^{-i\frac{2\pi}{3}y_0}$ again for all inputs. 

Before showing what is the maximal $\Delta$ achievable for such 
a wide class of classical protocols, we shall introduce its quantum competitor.
Let Alice and Bob  share a pair of entangled qutrits and suitable measuring devices (see, e.g. \cite{marek}).
This is their quantum protocol
\begin{itemize}
\item If Alice receives $x_0=0$, she
will measure her qutrit with the apparatus which is set to measure
a three-valued observable $A_0$. Otherwise, i.e. for $x_0=1$,
she sets her device to measure a different three-valued observable
$A_1$.
Bob follows the same protocol. If he receives $y_0=1$, he measures 
the three-valued observable $B_0$ on his qutrit. For $y_0=0$ he measures a different 
three-value observable $B_1$. We ascribe to the outcomes of the measurements the
three values $1$,
$e^{i\frac{2\pi}{3}}$ and $e^{-i\frac{2\pi}{3}}$ (the Bell numbers \cite{marek}).
The actual value obtained by Alice in the given measurement will be denoted again by $a$, whereas
the one of Bob's, also again, by $b$. 
\item Alice sends trit $e_A=y_1 \cdot a$ to Bob, and Bob sends trit $e_B=y_2 \cdot b $ to Alice.
\item Upon reception of the transmitted values they both broadcast $e_A\cdot e_B$ as their answers.
\end{itemize}

The task in both protocols is to maximize $\Delta=P(f_1)-P(f_2)$. The probability $P(f_1)$ is the probability
for the product $a \cdot b$ of the local measurement results to be 
equal to $e^{i\frac{2\pi}{3} (x_0 y_0)}$ in the two (classical and quantum) protocols:
\begin{eqnarray}
P({f_1}) &=& \frac{1}{4} [P_{A_0,B_1}(a b \!=\!1) +
P_{A_0,B_0}(a b\! =\!1)  \nonumber \\ & & + P_{A_1,B_1} (a b
\!=\!1) + P_{A_1,B_0} (a b\! = \!e^{i\frac{2\pi}{3}})],
\label{tigar}
\end{eqnarray}
where e.g. $P_{A_0,B_1}(a b \!=\!1)$ is the probability that $a
b =1$ if Alice measures $A_0$ and Bob measures $B_1$ (after she receives $x_0\!=\!0$
and he $y_0\!=\!0$). Recall that all four possible combinations
for $x_0$ and $y_0$ occur with the same probability 
$\frac{1}{4}$. Similarly the probability $P(f_2)$ that product  $a \cdot b$ is equal to
$e^{i\frac{2\pi}{3} (2\!-\! x_0\!-\! y_0)}$ is given by
\begin{eqnarray}
P(f_2) &=& \frac{1}{4} [P_{A_0,B_1}(a b \! =\!
e^{-i\frac{2\pi}{3}})\! +\! P_{A_0,B_0}(a b
\!=\!e^{i\frac{2\pi}{3}}) \nonumber
\\ & & + P_{A_1,B_1} (a b\!=\!e^{i\frac{2\pi}{3}}) \!+\! P_{A_1,B_0}
(a b \!=\! 1)].
\label{puma}
\end{eqnarray}
Finally, one notices that the success measure in the task is given by
\begin{equation}
\Delta=\frac{1}{4} I_3, \label{ris}
\end{equation}
where $I_3$ is exactly the Bell expression as defined by Collins {\it
et al.} \cite{collins}. It is the combination of probabilities obtained here when the right-hand side of
Eq. (\ref{tigar}) is subtracted by the right-hand side of Eq. (\ref{puma}) and then multiplied by
4. Collins  {\it et al.} \cite{collins} showed that $I_3 \leq 2$ for
all local realistic theories. Recently the violation of this inequality was demonstrated for a pair 
of spin-1 entangled photons \cite{ali}. In Ref. \cite{collins}
the local measurement results are defined differently (as numbers 0, 1, and 2); however
this description and the one used here are equivalent.

If one looks back at the family of classical protocols introduced above, one sees
that they are equivalent to a local realistic model of the quantum protocol ($\lambda$'s
are local hidden variables, and $x_1, y_1$ are some local variables which are not hidden).
This implies that within the full class of classical protocols considered here $\Delta \leq 0.5 $. 

\begin{figure}
\centerline{\psfig{width=9cm,file=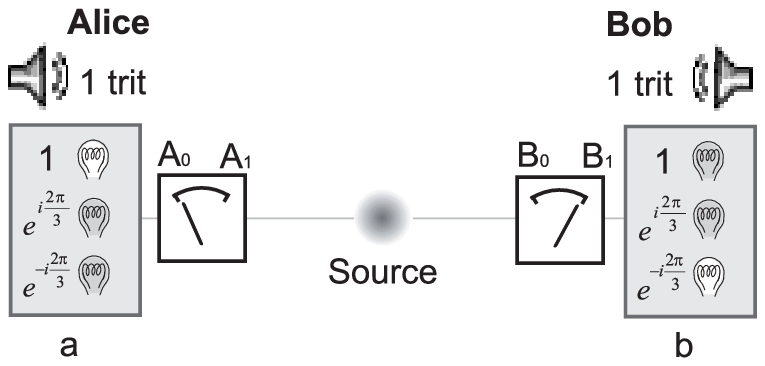}} \caption{Two-party quantum
communication complexity protocol which is based on the Bell-type experiment with entangled qutrits.
Alice receives a string $x = (x_0,x_1) $ and Bob $y = (y_0,y_1)$.
Depending on the value of $x_0$ Alice chooses to measure between two different
three-values observables $A_0$ and $A_1$. Similarly depending on $y_0$ Bob chooses to measure
between three-values observables $B_0$ and $B_1$. Alice's result of the measurement is
denoted by $a$ and Bob's by $b$. In the last step of the protocol
Alice sends the trit $y_1\cdot a$ to Bob and Bob sends the trit $y_2\cdot b$ to Alice.}
\label{bellexp}
\vspace{-0.5cm}
\end{figure}

Thus, the {\it necessary} and {\it sufficient}
condition for the state of two qutrits to improve the success in our communication complexity task over any
classical protocol of the discussed class is that the state violates the Bell
inequality for two qutrits. Note that, except for shared entanglement, the discussed classical and
quantum protocols are performed under the same conditions.
A wider class of classical protocols could include local calculations
of functions $a(x_0,x_1,\lambda_A)$, for Alice, and $b(y_0,y_1,\lambda_B)$, for Bob, which depend
on the full local inputs. Note that then the quantum competitor would be based on Bell's experiment
where the measurements can be chosen between six alternative three-valued observables. Since Bell's
inequalities for such situations are unknown, we restrict our consideration to the class of
(classical and quantum) protocols described above. However, the fact that $\Delta$ in the intuitive
classical protocols is equal to the maximal possible $\Delta$ in the discussed class of classical
protocols strongly indicates that this class, although not the most general one, might already include the
optimal one. This could be due to the different role of the entries
$x_0,y_0$ and $x_1,y_1$ in the functions $f_1$ and $f_2$.

It was shown in Ref. \cite{acin1} that a non-maximally (asymmetric) entangled state of
two qutrits that reads: $|\psi\rangle =
\frac{1}{\sqrt{2+\gamma^2}}(|00\rangle + \gamma|11\rangle +
|22\rangle)$ with $\gamma =(\sqrt{11}-\sqrt{3})/2 \simeq 0.7923 $
can violate the Collins {\it et al.} Bell inequality stronger than the
maximally entangled one. In that case the Bell expression $I_3$
reaches the value $ 1 + \sqrt{11/3} = 2.9149$. This implies that
with the use of this particular state the probability difference
$\Delta$ in our protocol can be as large as 0.729.

Therefore in a classical protocol, even with shared random
variables, more than 2 trits of information are {\it necessary} to
complete the task successfully with $\Delta$ at
least 0.729, whereas with the quantum entanglement 2 trits
are {\it sufficient} for the task with the same $\Delta$. Note
that the discrepancy between the measure of success in
the classical and the quantum protocol is higher here than in the
two-party communication complexity problem of Ref. \cite{buhrman} mentioned above.
Here we have $\Delta_Q \!- \!\Delta_C\! \approx \! 0.23$ whereas there
$P_Q\!-\!P_C\! \approx \!0.1$.

As another example we formulate a standard communication complexity task
which is an immediate generalization of the one of Ref. \cite{buhrman}. 
In this case the task of Alice and Bob is only to maximize the probability for
correct computation of function $f_1$. Since this is just a
first part of the task introduced above the highest possible probability of
success in a classical protocol is $P_C\!=\!0.75$.
The connection with the violation of a Bell's inequality is established through equation
$P_C\!=\! I_2/4$, where $I_2$ is another Bell expression (which is equal to Eq. (\ref{tigar}) with the factor $\frac{1}{4}$ 
dropped)
introduced in Ref. \cite{collins}. For all local realistic theories $I_2\!\leq \!3$.
Therefore all quantum states of entangled qutrits which violate this Bell inequality can lead
to higher then classical success rate for the task.

We note that a series of similar specific two-party communication
complexity tasks can be formulated
with quantum solutions which exploit the possibility
of two arbitrarily high-dimensional quantum systems to violate the
corresponding Bell inequalities of Ref. \cite{collins}.

As noted in Ref. \cite{gisin} one may ask whether the use of qutrits is necessary for
any quantum information task, because qutrits can be teleported with help of singlets and classical
communication. Yet any such realization would require more communication than permitted by our
protocol. One may also ask whether the exclusive use of the states which violate Bell's inequalities
is necessary for the problem, as there are
non-separable states which do not directly violate Bell's inequalities but only after local operations and
communication \cite{popescu}. Yet again such transformation would require additional communication.

We interpret our work as a further example
suggesting that the violation of Bell inequalities can be considered as a
"witness of useful entanglement". This was first coined and suggested in
\cite{scarani,acin} in different contexts.

M.\.{Z}. acknowledges KBN grant No. 5 P03B 088 20. {\v C}.B. is
supported by the Austrian  FWF project F1506, and by the QIPC
program of the EU. The work is a part of the Austrian-Polish
program "Quantum Communication and Quantum Information IV" (2002-2003).

\end{document}